# Impact of geometry on light collection efficiency of scintillation detectors for cryogenic rare event searches

F.A. Danevich[1], V.V. Kobychev[1], R.V. Kobychev[1,2], H. Kraus[3], V.B. Mikhailik[*,3,4,]
V.M. Mokina[1] and I.M. Solsky[5]

[1] Institute for Nuclear Research, MSP 03680, Kyiv, Ukraine
[2] National Technical University of Ukraine "Kyiv Polytechnic Institute", 03056 Kyiv, Ukraine
[3] Department of Physics, University of Oxford, Oxford, OX1 3RH, UK
[4] Diamond Light Source, Didcot, OX11 0DE, UK
[5] Scientific Research Company CARAT, 79031 Lviv, Ukraine

**Abstract.** Simulations of photon propagation in scintillation detectors were performed with the aim to find the optimal scintillator geometry, surface treatment, and shape of external reflector in order to achieve maximum light collection efficiency for detector configurations that avoid direct optical coupling, a situation that is commonly found in cryogenic scintillating bolometers in experimental searches for double beta decay and dark matter. To evaluate the light collection efficiency of various geometrical configurations we used the ZEMAX ray-tracing software. It was found that scintillators in the shape of a triangular prism with an external mirror shaped as truncated cone gives the highest light collection efficiency. The results of the simulations were confirmed by carrying out measurements of the light collection efficiencies of $CaWO_4$ crystal scintillators. A comparison of simulated and measured values of light output shows good agreement.

## 1. Introduction

Scintillation detectors have a wide range of applications in medicine, industry, security and scientific research. In many of these applications they are required to have good energy resolution, and that can be achieved by maximising the number of photons reaching the actual photodetector. Therefore, the improvement of light collection efficiency (ratio of photons arriving at the photodetector to the total number of photons generated in the scintillator) is and has always been a crucial component in the optimization of scintillation detectors. This parameter is of high importance for a special kind of cryogenic phonon-scintillation detectors in which scintillation and phonon signals are induced in a crystal by interaction of a high-energy particle or photon [1]. These detectors combine exceptional energy resolution (reading out the phonon channel signal) with the ability to distinguish the type of interaction event (charged particles, gamma quanta or neutrons) using the response from the scintillation channel [2]. Such features make them the detectors of choice in the experimental search for rare events where the sensitivity of the experiment relies on a low energy threshold and good event type discrimination ability of the detector [3]. Maximising the amount of detected light allows improving these characteristics of cryogenic phonon-scintillation detectors. Furthermore, improvement of the signal-to-noise ratio by optimization of light collection plays a crucial role in the discrimination of randomly coinciding events in cryogenic scintillating bolometers. For instance, coincidence of two neutrino double beta decay events could be one of the main sources of background for neutrinoless double beta decay in cryogenic bolometers due to their poor time resolution [4]. Therefore, the optimization of light collection efficiency of a scintillation detector is an important task for the next generation of cryogenic experiments searching for rare events.

A significant fraction of photons generated in a scintillator remains trapped in the volume due to total internal reflection and this contributes to losses by self-absorption [5]. There are several methods that allow tackling this issue. Coupling the crystal to the photodetector with optical grease leads to better light collection through increasing the critical angle of total internal reflection [6]. Application of reflective wrapping allows redirecting photons towards the detector, which are normally lost as they exit through the sides [5], [7], [8]. The purpose of surface treatment (grinding, lapping, etching or painting) is similar to wrapping, i.e. aiming to backscatter the light into the crystal [9], [10].

---

[*] Corresponding authors e-mail address: vmikhai@hotmail.com

Modification of the scintillation crystal's geometry is up to now the least explored option due to practical constraints and considerations. Nonetheless, there are a few studies of the geometry effect, showing, for example that light collection improves systematically with increase of aspect ratio (width-to-height) of a scintillator [9], [11], [12], [13], [14]. Breaking the symmetry of a scintillation crystal was also found to be advantageous for high light output [15]. In line with these findings, recent Monte Carlo studies of the effect of crystal shape indicated that a noticeable improvement of the light response of the detector can be achieved by reducing the symmetry of the scintillator crystal [16].

There are a number of constrains for the application of these methods in the case of cryogenic experiments where the same crystal is used as a scintillator and a bolometer. The scintillation crystal must be separated as best as possible from the rest of the detector through a gap (vacuum) to minimise phonon losses and to avoid excess heat capacity. This is incompatible with optical coupling or wrapping. Therefore, the improvement can be gained from optimizing type of surface treatment and the optimum shape of the scintillation detector. There is a lack of published results on the effect of crystal geometry on the light collection efficiency of scintillation detector used in cryogenic experiments. Only recently, we studied the effect of the crystal shape and surface treatment on the light collection efficiency of a $ZnWO_4$ scintillation detector and demonstrated that light collection efficiency and energy resolution of a scintillator in the shape of a hexagonal prism is improved when compared with a cylindrical scintillator [17]. This paper contributes results and findings on the relationship between light collection efficiency and geometry of scintillation detector used in cryogenic rare event search experiments. In this work we first investigate the effect of different modifications theoretically with the aid of Monte Carlo simulations and after that we verify the predictions experimentally. This letter reports on an finding emerging from this investigation that will inform the design of scintillation detection modules for cryogenic rare event search experiments.

## 2. Model and simulations

For this study we selected a $CaWO_4$ scintillator that is being used in cryogenic experiments for more than a decade [18], [19]. The crystal has high scintillation light yield at cryogenic temperature and negligible anisotropy. We simulated the geometry of this scintillation detector in a configuration that is relevant for cryogenic experiments: we model a setup in which the crystal is surrounded by an external specular reflector with a gap 3 mm and there is a gap of 0.3 mm between the crystal and the photodetector (see Fig. 1). Our aim was to investigate the effect of scintillation detector geometry on light collection efficiency, and thus we extended building on results of [17] the choice of shapes by experimenting with cylinder, hexagonal, rectangular and triangular prisms. The scintillation elements are of the same height $h$ and their bases are inscribed in a circle of fixed diameter $d$. The values of $h$ and $d$ correspond to those of our experimental samples (see section 4). We also investigated how light collection varies with the shape of the external reflector where this is either a cylinder (R1) or a truncated cone (R2). Finally, we considered two cases for the crystal surface a) all polished and b) side surface of the scintillation crystal diffuse with face surfaces polished.

The modelling of light transport in the scintillation detector was done using the ZEMAX ray-tracing software [20]. The simulations were carried out in non-sequential mode, where rays are traced through the objects without predefined sequence of surfaces that traced rays must impinge upon, i.e. they may hit any part of any object and also they may hit the same object multiple times. The trajectories of rays are determined solely by the physical positions and properties of the objects and directions of the rays. This allows accounting for the total internal reflections as well as taking into consideration polarisation effects and ray splitting at the interfaces.

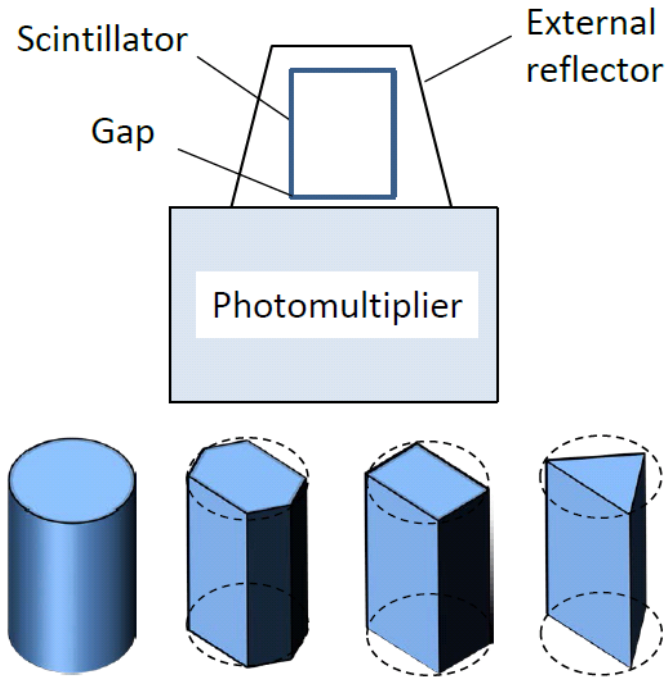

Fig. 1. Schematic of experimental setup showing scintillation crystal surrounded by external specular reflector without optical contact to the photodetector (upper part). Four different shapes of scintillation crystals were used in simulations (lower part).

The scintillation source was simulated as a volume object of given shape with dimensions 0.001 mm less than the geometrical sizes of scintillation crystal to eliminate uncertainty at the surface. The scintillation photons are assumed to be isotropically distributed and not polarised. The wavelength was set to the peak of the $CaWO_4$ emission at 420 nm [1]. The simulations where done in isotropic approximation using an average dispersion curve and refractive index n =1.96 at 420 nm as the small anisotropy of calcium tungstate ($n_e-n_o$=0.02) has negligible effect on the final results. The absorption coefficient of the crystal $\alpha$=0.11 $cm^{-1}$ was obtained from measured transmission spectra corrected for multiple reflections following [21]. The value of the bulk scattering coefficient of $CaWO_4$, $\sigma$= 0.06 $cm^{-1}$ was determined in previous studies of this scintillator [22]. The entrance window of the photomultiplier (PMT) was modelled as a 1 mm thick disk, 76 mm in diameter, made of borosilicate glass BK7 with refractive index n=1.52. For the calculation of light collection efficiency the detector was assumed to be an ideal absorber at the inner surface of the PMT window. In experiment, the condition of side surface of the crystals was subsequently modified by lapping. Such treatment changes the surface state from polished to diffused, and that was modelled as a Lambertian reflector.

In each case, 100,000 photons (each at a fixed wavelength of 420 nm) randomly distributed over the volume of the crystal were traced. The statistical error of the simulations is ±0.3 %. The result of the simulations is the fraction of the total energy generated by the source object (scintillator) that reaches the detector. This number represents the light collection efficiency of the setting under study.

## 3. Results of simulations

Using the input parameters and model discussed above, the light collection efficiency for the different configurations was studied through simulation. The results of modelling presented in Fig. 2 exhibit two slightly different trends depending on the surface treatment. For polished crystals the light collection efficiency increases from cylinder to hexagonal prism then reduces a little for the rectangular prism and finally reaches a maximum for the triangular prism, a behaviour that is confirmed experimentally (see Sec. 4). It should be noted that according to recent studies [16] the amount of light escaping the crystal through the side surfaces depends on the angle between the surfaces, leading to minima and maxima at certain angles. It is very likely that the behaviour observed here is the manifestation of this effect.

In the case of diffuse side surfaces the trend is monotonic: the light collection efficiency shows a monotonic increase as the shape of the scintillation crystal changes from cylinder to hexagonal prism and rectangular prism finally reaching a maximum for the triangular prism. The ratio of the light collection efficiency of triangular prism to that of cylinder is about 1.5 irrespectively of surface treatment.

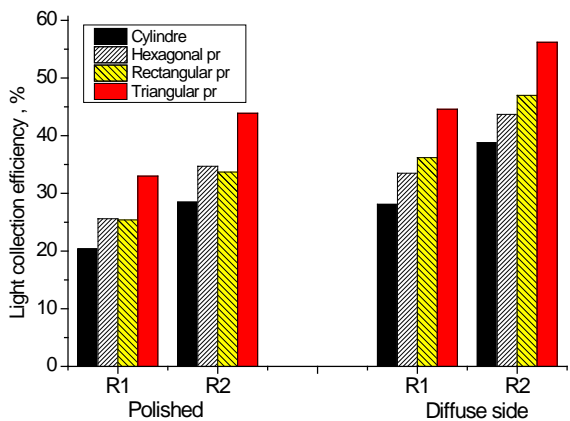

Fig.2. Simulated values of light collection efficiency for $CaWO_4$ scintillators of different shape in the conditions without optical contact between scintillator and photo-detector. The annotation for the horizontal axis identifies the type of external reflector (R1: cylinder or R2: truncated cone) and type of surface treatment.

Furthermore, there is a clear dependence on the geometry of the external reflector. In all cases the reflector in the shape of a truncated cone is superior when compared to a cylinder. The inclination of the reflective surface allows redirecting more photons in the gap away from the crystal, thereby enhancing their chance to reach the photo-detector. Lapping of the surfaces of the scintillation crystals brings about another improvement so that the simulated value for the light collection efficiency of the triangular prism in the configuration with truncated cone reflector (R2) exhibits a very high value of 57%. It is worth noting that the simulated values of light collection for the diffused cylindrical and hexagonal scintillators with cylindrical reflector are equal to 30 and 35%, respectively. This is very close to the results of our Monte Carlo simulations reported for $ZnWO_4$ scintillators in the same setting [17]. In section 4 we present results on experimental verification of these finding.

## 4. Experimental verification

To verify the predictions obtained through modelling, four scintillation elements of different shape (with $d = 17.6$ mm and $h = 14.6$ mm, see section 2) were produced from one $CaWO_4$ crystal. This was to ensure that the optical properties of the samples are as similar as possible. The measurements were carried using a 3" PMT Philips XP2412. To ensure identical optical gap between scintillator and PMT window for all settings we placed the crystal on two glass fibres with diameter 0.3 mm, each. A small ($5\times5\times1$ mm$^3$) $CaWO_4$ scintillator with a collimated $^{241}$Am α-source was optically coupled to the PMT window and was used to monitor the stability of the setup. This allows accounting for any changes during measurements and it improved the overall accuracy of the experiment. We have estimated the error of the peak position measurements as ±2%. The experiment consisted of recording energy spectra of $^{137}$Cs and $^{207}$Bi sources from polished $CaWO_4$ crystal scintillators using two external reflectors (cylinder or truncated cone) and repeating these experiments for diffuse side surfaces.

Fig. 3 shows an example of a $^{137}$Cs energy spectrum recorded for the triangular prism of a $CaWO_4$ scintillator with diffused side and polished end faces, surrounded by a reflector in the shape of a truncated cone. It is noticeable that inspite of the absence of optical coupling in this configuration, the measured energy resolution (full width at half maximum, FWHM) 6.36% is the best ever reported for a $CaWO_4$ scintillator (compare with 7.2% [23], and 6.6% [24]).

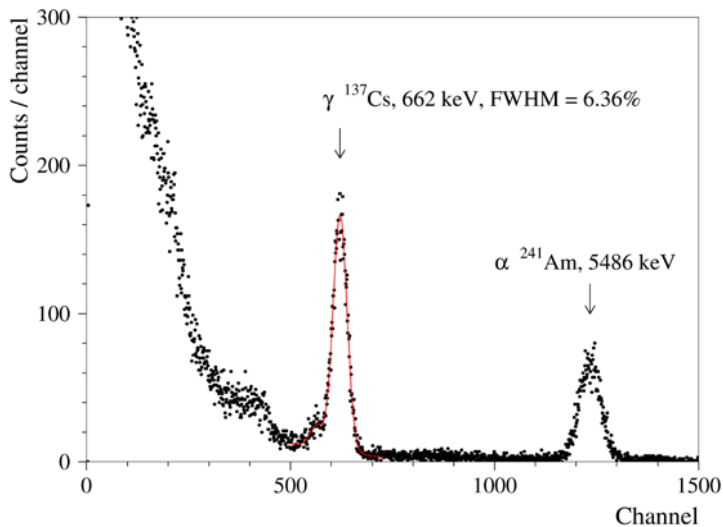

Fig. 3. Energy spectrum of $^{137}$Cs γ quanta measured for the triangular prism $CaWO_4$ crystal scintillator with diffused side and polished end faces without optical contact with the PMT surrounded by external truncated cone reflector. The peak at channel 1230 is the alpha peak of $^{241}$Am installed on the "reference" $CaWO_4$ crystal $5\times5\times1$ mm$^3$ used to stabilize the spectrometer gain.

The summary of measurements of energy resolutions with 662 keV γ quanta of $^{137}$Cs for different experimental configurations is presented in Fig. 4. The striking feature of this plot is the poor energy resolution of the polished cylindrical scintillator. The same effect is observed in [17] where it is explained by non-uniformity of light collection from the cylinder. In this case, the polished cylinder behaves as a light guide, exhibiting significant radial variation of the light distribution [25], which is the cause of degradation in energy resolution. Lapping of the cylinder's surfaces breaks total internal reflection, alleviates the non-uniformity of light collection and improves energy resolution.

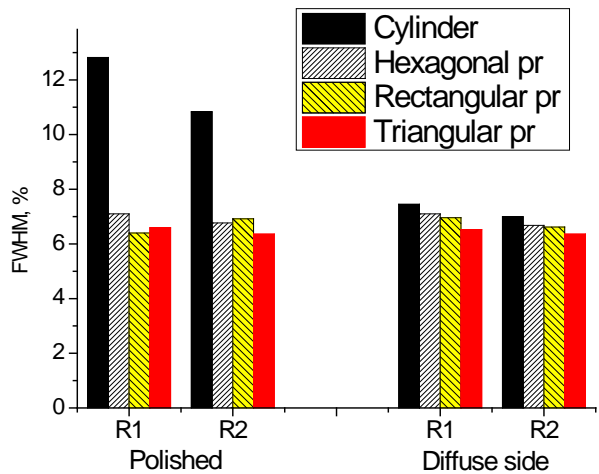

Fig. 4. Energy resolutions for cylinder, hexagonal, rectangular and triangular prisms of a CaWO$_4$ scintillator, measured with γ from a $^{137}$Cs source for different properties of the crystal surface and external reflectors without optical contact between scintillator and photodetector.

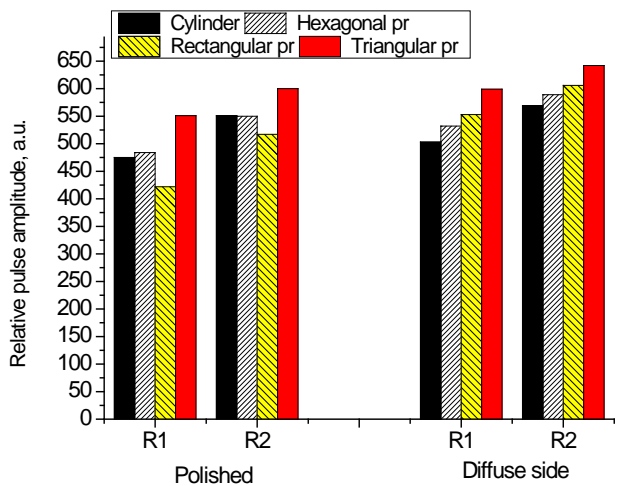

Fig. 5. Relative pulse amplitude for cylinder, hexagonal, rectangular and triangular prisms of a CaWO$_4$ scintillator crystals measured with γ from a $^{137}$Cs source for different properties of the crystal surface and external reflectors without optical contact between the scintillator and the PMT.

The relative pulse amplitudes measured with 662 keV γ quanta of $^{137}$Cs for different conditions of experiments are presented in Fig. 5. Juxtaposing these results with the data in Fig. 4 we can see good correlation between the increase of pulse amplitude and improvement of energy resolution (with the exception of polished cylinder as explained above). The correlation is particularly clear in the case of scintillation crystals with diffuse side surfaces. In general, there is steady increase of the relative pulse amplitude and subsequent improvement of energy resolution as shape of the scintillation crystal changes from cylinder to hexagonal prism then rectangular prism and finally triangular prism

The main interest of our study is comparison of these experimental findings with the results of simulations shown in Fig. 2. Comparing the simulated data in Fig. 2 with the experimentally obtained data in Fig. 5, it is clear that qualitatively our simulations reproduce well the main relationships between the results. That is to say that the predicted variation of the light collection efficiency with the crystal shape is consistent with that observed by measuring the change of relative pulse amplitude. The general tendency in the effect of the crystal shape on the light collection efficiency is also consistent with the model: for all examined configurations the triangular prism exhibits the highest light collection. The maximum light output is detected for the scintillation detector in the shape of triangular prism with diffused side surfaces and truncated conical reflector.

There is also consistency with reproducing irregularities in the general behaviour of the light collection efficiency. For instance, for the case of a polished scintillator, the detected relative pulse amplitude for the rectangular prism is lower than that of the hexagonal prism, being consistent with the prediction. However at this point there are certain quantitative discrepancies between the model and experiment. More specifically, the measured improvement of light collection efficiency is rather modest as compared to what is predicted by simulations. The observed increase of the light output for the triangular prism in comparison with that of the cylinder is about 20% while the predicted value is 50%. There are a few factors that can contribute to this difference. Firstly, the photon scattering at the edges [11] that facilitates extraction of photons from the crystal is not implemented in our model. Secondly, there is a possible discrepancy in the modelling of the diffuse surface, which is very difficult to simulate [10]. We considered the diffuse crystal surface in Lambertian approximation, thus any contribution from a specular component is ignored in the model.

## 6. Conclusion

We presented in this paper the results of a study aiming to demonstrate the effect of crystal shape and detector geometry upon the light collection efficiency of scintillation detectors in configurations without optical coupling, a situation most relevant for cryogenic experiments. For this we employed a theoretical model developed recently for simulation of light collection efficiency of scintillation detection modules [17]. First, we calculated the light collection efficiency of $CaWO_4$ scintillator in the shape of cylinder, hexagonal, rectangular or triangular prism with two different external reflectors, i.e. cylindrical and truncated cone. The simulation studies indicated that a triangular prism with diffuse side surface surrounded by a truncated conical reflector provides the highest light collection efficiency.

For experimental verification of the simulations, crystal scintillators of shapes used in simulations where produced from a single $CaWO_4$ sample and their scintillation characteristics were measured for the different configurations. We observed very good agreement between the simulation and experimental results; all major trends and model predictions are supported by the experimental findings. It is concluded that from the viewpoint of light collection efficiency the best configuration for a cryogenic scintillation detection module is achieved when the crystal is shaped as triangular prism with diffused side surface and surrounded by an external reflector in the shape of a truncated cone, in agreement with our simulations. Because of the improvements of light collection efficiency in this configuration with air gap between scintillator and photodetector it was possible to achieve an excellent energy resolution of 6.36% - a value which has been so far attainable only by using optical coupling. This configuration is therefore recommended for cryogenic scintillation detector used in the search for rare events.

**Acknowledgments**

The study was supported in part by a grant from the Royal Society (London) ''Cryogenic scintillating bolometers for priority experiments in particle physics''. The group from the Institute for Nuclear Research (Kyiv, Ukraine) was supported in part by the Space Research Program of the National Academy of Sciences of Ukraine. Thanks to Mr. Kudovbenko for technical support with machining components of experimental setup.